# The Trial within Cohorts (TwiCs) study design in oncology: Experience and methodological reflections.

Rob Kessels[1], Anne M. May[2*], Miriam Koopman[3], Kit C.B. Roes[4]

[1] Dutch Oncology Research Platform, Netherlands Cancer Institute, Amsterdam, the Netherlands.
[2] Julius Center for Health Sciences and Primary Care, University Medical Centre Utrecht, Utrecht University, Utrecht, the Netherlands.
[3] Department of Medical Oncology, University Medical Centre Utrecht, Utrecht University, Utrecht, the Netherlands.
[4] Radboud University Medical Center, Department of Health Evidence, Section Biostatistics, Nijmegen, the Netherlands.

[*]Correspondence:
A.M.May@umcutrecht.nl
Julius Center for Health Sciences and Primary Care, University Medical Center Utrecht, Utrecht University, Utrecht, the Netherlands
Postal address STR 6.131 | P.O. Box 85500| 3508 GA Utrecht

**Abstract**
A Trial within Cohorts (TwiCs) study design is a trial design that uses the infrastructure of an observational cohort study to initiate a randomized trial. Upon cohort enrollment, cohort participants provided consent for being randomized in future studies without being informed. Once a new treatment is available, eligible cohort participants are randomly assigned to this new treatment or standard of care. Patients randomized to the treatment arm are offered this new treatment, which they can refuse. Patients who refuse will receive standard of care instead. Patients randomized to the standard of care arm, receive no information about the trial and continue receiving standard of care as part the cohort study. Standard cohort measures are used for outcome comparisons. The TwiCs study design aims to overcome some issues encountered in standard Randomized Controlled Trials (RCTs). An example of an issue in standard RCTs is the slow patient accrual, which a TwiCs study aims to improve by selecting patients using a cohort and only offering the intervention to patients in the active arm. In oncology, the TwiCs study design has gained increasing interest during the last decade. Despite its potential advantages over RCTs, the TwiCs study design is concerned with several methodological challenges that need careful consideration when planning a TwiCs study. In this article, we focus on these methodological challenges and reflect on them using experiences from TwiCs studies initiated in oncology. Important methodological challenges that are discussed are the timing of randomization, the issue of non-compliance (refusal) after randomization in the intervention arm, and the definition of the intention-to-treat effect in a TwiCs study and how this effect is related to its counterpart in standard RCTs.

**Introduction**

Randomized Controlled Trials (RCTs) are generally considered the gold standard in experimental design for evaluating the efficacy of medical treatments. By randomly allocating patients to different treatment groups, it is expected that prognostically comparable groups are created that only differ in the assigned treatment and hence, different sources of bias are minimized. The standard approach is typically to conduct randomized clinical trials each investigating the effect of a single intervention in a single disease. However, it has been argued that this 'classical' way of conducting clinical research has become challenging due to several reasons, including (but not limited to) uncompleted trials, high drop-out rate in the control group and the limited external validity. These three reasons will be discussed below.

First, it has been argued that standard RCTs in oncology often remain uncompleted because they are confronted with poor and slow accrual of patients, high costs, and inadequate funding [1]. Slow accrual might be due to the fact that only a small fraction of all cancer patients are actually enrolled in a trial, because, for example, patients refuse to be randomized. Also, due to an increased heterogeneity in tumor types, the number of small subgroups increases. This leads to trials where only a small subset of patients is eligible, making accrual even more challenging.

Second, the issue of high drop-out in the control group might be caused by disappointment of patients who are randomized to a control treatment [2]. This is especially true for open-label trials that cannot be performed in a blinded setting. Many oncology RCTs are more likely to be open-label trials [3] due to differences in the administration of treatments or differences in treatment schemes between treatment arms. For example, comparing two different radiotherapy schemes cannot be performed in a blinded setting. As a consequence, one could expect differential drop-out in the groups.

Third, standard RCTs have been criticized of their limited external validity, because participants selected in RCTs are generally different compared to the population [4]. Clinical trials tend to exclude elderly patients and/or patients with common comorbidities [5] and phase 3 clinical trials often fall short of including a representative number of patients from diverse racial and ethnic groups [6]. Also, RCTs generally implement many eligibility criteria which diverge from the traditional disease definition. Hence, RCTs are prone to selection bias. External validity is also affected if the RCT does not mimic routine clinical practice, because in clinical practice, for example, patients are not informed about an (complete) overview of possible experimental therapies. Furthermore, participating in a trial often involves more frequent and closer monitoring of patients compared to routine clinical practice and this might lead to different results observed in trials compared to clinical practice. For example, progression of disease might be detected earlier in trials or patients report more positively about their quality of life during trial participation due to the close attention that is given to them by trial staff. These external validity issues lead to concerns regarding the generalizability of results. Pragmatic trials aim to (partly) solve these external validity issues by retaining the randomization component and adopting characteristics of routine clinical practice [7]. However, as pragmatic RCTs are still 'classic' RCTs, they may still face potential disadvantages like poor accrual, and risk of drop-out.

The abovementioned problems of standard RCTs raised interest for methodological innovation. The Trial within Cohorts (TwiCs) study design can be regarded as such an innovation and was developed to address some of the issues encountered in standard RCTs. The TwiCs design was originally proposed by Relton et al. [8], who introduced it as the 'cohort multiple randomized controlled trial'. A TwiCs study is performed within an observational cohort study following patients diagnosed with the condition of interest and who receive Standard of Care (SOC). At cohort enrollment, patients have given informed consent for cohort participation after which data on clinical and for example patient reported outcomes are regularly collected at baseline and follow-up. Additionally, they provide consent for being randomized in future studies without notice. Once an alternative treatment becomes available, a clinical trial will be designed that is conducted within the cohort to evaluate the effectiveness and safety of this alternative treatment. In this trial, the SOC (that all cohort patients are already receiving) will serve as the control treatment. Within the cohort sample, eligible patients are identified, and these patients are randomly assigned to the alternative treatment or SOC. Patients randomized to the alternative treatment arm are requested to provide additional informed consent and are then offered the alternative treatment. Patients randomized to SOC are not informed about the trial, but they take part in the trial as controls, making use of the standard follow-up of the observational cohort. Furthermore, patients randomized to the alternative treatment can decline this treatment after which they will receive SOC as well. The rationale for the TwiCs design is that it prevents (possible) drop-out and contamination in the control group as these patients are not informed about the trial and therefore cannot get disappointed about the fact that they do not receive a new alternative treatment. Also, because a TwiCs study uses the infrastructure of an observational cohort study, it can improve accrual rate and accrual speed since eligible patients are already known and can therefore be approached easily. In addition, the availability of an observational cohort infrastructure makes it possible to base eligibility on information from routine cohort measurements. When adopting such a selection procedure, it may ensure that selection of patients in a TwiCs study is less affected by the preference of a physician and hence, selection bias will be reduced [9]. Furthermore, a TwiCs study aims to mimic clinical practice as the control group receives SOC and is assumed to be unaware of an experimental treatment that they cannot receive. Reduction of selection bias and/or mimicking clinical practice improves representativeness and the generalizability of the trial results. However, it is crucial that the follow-up schedule of the trial does not diverge from the follow-up schedule of SOC as otherwise outcome measures will not be available. In other words, it is important that the SOC follow-up schedule matches the purpose and the research question of the TwiCs study, as it is undesirable or even impossible to adapt the SOC follow-up schedule.

      Because a TwiCs study is performed within an observational cohort study, it brings in another possible advantage: to perform multiple RCTs over time using patients of the same cohort (from where it owes its original name 'cohort *multiple* RCT'). This enables the investigation of multiple treatments within the same patient group. This capacity is aligned with the recent development of master protocols to study multiple therapies for a single disease, a single therapy for multiple diseases or both [10], which have been initiated for the study of cancer therapies, see for example [11]. Woodcock and LaVange [10] discuss three master protocol studies: the umbrella trial, the basket trial and the platform trial. In an umbrella trial, multiple therapies are tested in the context of a single disease, whereas in a basket trial, a single therapy is tested in the context of multiple diseases. In a platform trial, multiple therapies are studied in the context of a single disease in a perpetual manner: therapies are allowed to leave or enter the platform based on a decision algorithm. As with the TwiCs design, master protocol studies were also developed to face the increasing costs and other challenges of standard RCTs, such as poor accrual due to precision medicines that are only suited for a small subset of patients. The availability of an observational cohort where enrolled patients are all diagnosed with the same disease enables the execution of an umbrella or platform trial using a master protocol. A TwiCs study can be placed in the context of a platform trial if

multiple TwiCs studies are performed to study the effect of multiple alternative treatments for treating the disease of interest.

Given these assumed advantages, TwiCs has gained interest in several research fields such as psychosocial and rehabilitation interventions [12], interventions for mental illness [13], and depression treated by homeopaths [14], amongst others. In this article, we will focus on the application of TwiCs designs within the oncology setting. Specifically, the TwiCs design is entailed with several methodological challenges and considerations of which the implications with respect to the applicability of TwiCs studies for (future) oncology trials will be discussed, using experiences from previous TwiCs studies in oncology. In this discussion, we will pay attention to the question whether a TwiCs study is able to answer the same research question that is usually answered by performing a standard RCT.

In addition to methodological challenges, the TwiCs design also faces several ethical issues [15,16]. Because the focus of this article is on the methodological challenges, these ethical issues will not be discussed here. However, we would like to point out one important ethical aspect that is concerned with randomizing patients without prior consent and keeping control patients oblivious to trial participation. Initially, only patients randomized to the alternative treatment receive informed consent in order to more closely resemble the informed consent procedure used in routine clinical practice [8]. In clinical practice, patients are usually not told about treatments they cannot receive. Although it more closely resembles clinical practice, this procedure introduces ethical controversies in the clinical trial setting. To manage these controversies, Young-Afat et al. proposed the staged-informed consent procedure [17]. The staged-informed consent procedure works as follows. At cohort enrollment, in addition to giving informed consent about cohort participation, patients are furthermore asked informed consent to be randomized in future RCTs within the cohort. Cohort participants are informed that they will be offered an alternative treatment if they are randomly assigned to this alternative treatment. They are also informed that they will otherwise serve as controls without being notified and that their data will be used to evaluate the alternative treatment. Subsequently, when a RCT is conducted within the cohort, eligible patients are selected and one of the eligibility criteria is whether cohort participants gave informed consent to be randomized in future RCTs. Of these eligible participants, a random selection will be invited to undergo the new treatment or intervention and these participants receive a new informed consent providing them all information of this new, alternative treatment. Participants who are not randomly selected receive SOC, are not informed about the trial and their cohort data is used to evaluate the effectiveness of the new intervention. At the end of trial, a third stage was introduced in which all cohort participants, irrespective of specific trial participation, receive the trial results (e.g. in an annual newsletter). The staged-informed consent overcomes that participants are randomized without their prior consent which mitigates ethical concerns, but caution is warranted since ethics committees may vary in how they interpret this [15]. The understanding of the staged-informed consent procedure was evaluated among patients participating in oncology TwiCs studies and it was found that patients did not have ethical objections to serve as control without further notice [18,19].

The article is structured as follows: in the next section, we provide an overview of known TwiCs studies within the oncology setting. Then, we discuss the methodological considerations faced in TwiCs studies and explain if and how these considerations were dealt with in the oncology TwiCs studies. This article ends with a discussion concerning future considerations.

**Main text**

**Overview of (applied) TwiCs in the oncological setting**

Within the oncology setting, there are several examples of initiated TwiCs studies. These studies are described in Table 1. All trials were or are still conducted within The Netherlands, with exception of the TILT trial, which was conducted in the UK. Four trials (TILT, RECTAL BOOST, UMBRELLA FIT, and VERTICAL) have now been completed while the other three trials are still ongoing at this moment. The TILT trial was the only (and first) trial that used a TwiCs design to study the effect of an investigational medicinal product. The TILT trial was a feasibility study to verify the feasibility of performing a randomized trial of intra-pleural bacterial immunotherapy using a TwiCs design with the pre-specified recruitment, attrition, and data completeness as primary outcome measures.

In all trials, the experimental group is offered an alternative treatment/intervention while the control group receives SOC. In the MEDOCC-CrEATE trial, a somewhat different study procedure is undertaken. The MEDOCC-CrEATE trial is conducted within the Prospective Dutch ColoRectal Cancer Cohort (PLCRC), an observational cohort study for patients diagnosed with colorectal cancer. The MEDOCC-CrEATE trial investigates the willingness of patients to receive adjuvant chemotherapy after detection of circulating tumor DNA and to assess the effect of circulating tumor DNA guided adjuvant chemotherapy. More specifically, about one week after surgery, eligible patients (who also provided consent to be randomized in future trials) for the MEDOCC-CrEATE trial are randomized to the intervention or control arm. Following the TwiCs study design, only patients randomized to the intervention arm are asked informed consent for the immediate analysis of their circulating tumor DNA status of a post-surgery blood sample. Patients who have detectable circulating tumor DNA in their blood, are offered adjuvant chemotherapy, which patients can either accept or refuse. Refusing patients receive control treatment, which consists of routine post-surgery follow-up care. Also patients without detectable circulating tumor DNA in their blood (as well as patients who provide no informed consent for the immediate analysis of circulating tumor DNA) receive control treatment. Patients randomized to the control group receive no information about the trial and their post-surgery blood samples are also not immediately tested for circulating tumor DNA. The difference with regular TwiCs studies is that in the MEDOCC-CrEATE trial, patients are not directly randomized to be offered alternative treatment, but they are primarily randomized to the possibility to find out whether they have circulating tumor DNA in their blood, which will then determine if they are offered an alternative treatment.

In addition to the trials and related cohorts listed in Table 1, more cohorts have been set-up where the TwiCs design has been introduced. For example, in the Netherlands, cohort studies for bladder cancer [20], gastro-intestinal cancer [21], and pancreatic cancer [22] have been initiated in which TwiCs studies can be embedded and also in the UK, a bladder cancer cohort has been initiated [23]. Furthermore, in prostate cancer, starting an observational cohort study has been proposed to build an infrastructure where multiple TwiCs studies can be performed to solve recruitment issues observed in prostate cancer RCTs [24]. The authors first started a pilot study to verify if patients are willing to participate in a cohort study and what they think of the staged informed consent, amongst other things [24]. In addition, in the prostate cancer setting, the TwiCs design was mentioned as promising trial design to solve recruitment issues when comparing focal therapy to active surveillance, radical therapy, or prostatectomy in a randomized setting [25,26,27,28].

**Table 1. Overview of applied TwiCs studies within the oncological setting**

| Trial | Description | Status | Planned number of patients |
|---|---|---|---|
| TILT [29,30] | Feasibility trial of an investigational medicinal product to treat mesothelioma. The aim of this study was to answer the question whether a full-scale version of the trial is possible and whether a TwiCs study is appropriate for mesothelioma trials. The investigational product is called OK-432 ("dead" bacteria) which is used to stimulate immune cells to attack the mesothelioma. The trial was initiated within the ASSESS-meso cohort [31]. | Completed | 45 patients |
| RECTAL BOOST [32,33,34] | Randomized controlled trial for pre-operative dose-escalation BOOST in locally advanced rectal cancer. The trial was originally initiated within the prospective data collection initiative on colorectal cancer (PICNIC) cohort, which has been renamed to the prospective data-collection initiative on colorectal cancer, Prospective ColoRectal Cancer Cohort (PLCRC) [35,36]. | Completed | 120 patients |
| HONEY [37] | Clinical trial of assessing the effect of hyperbaric oxygen therapy in breast cancer patients with late radiation toxicity. The trial is initiated within the Utrecht cohort for multiple breast cancer intervention studies and long-term evaluation (UMBRELLA)" [38]. | Ongoing | 120 patients |
| UMBRELLA FIT [39,9,40] | Clinical trial investigating the effect of an exercise program on the quality of life of patients with breast cancer. The trial was initiated within the UMBRELLA cohort [38]. | Completed | 192 patients |
| MEDOCC-CrEATE [41] | Clinical trial investigating the effect of circulating tumor DNA guided adjuvant chemotherapy in stage II colon cancer. The presence of circulating tumor DNA is only assessed in the alternative treatment group. The trial is initiated within the PLCRC cohort [35,36]. | Ongoing | 60 patients with circulating tumor DNA. In total, 1320 patients. |
| SPONGE [42] | Clinical trial investigating the impact of retractor SPONGE-assisted laparoscopic surgery on duration of hospital stay and postoperative complications in patients with colorectal cancer. The trial is initiated within the PLCRC cohort [35,36] and is a follow-up trial of the RECTAL BOOST trial: patients of the RECTAL BOOST trial were also eligible for the SPONGE trial. | Ongoing | 196 patients |
| VERTICAL [43,44,45,46] | Clinical trial comparing conventional radiotherapy with stereotactic body radiotherapy in patients with spinal metastases. The trial is initiated within the prospective evaluation of interventional studies on bone metastases (PRESENT) cohort [47]. | Completed | 110 patients |

## Methodological considerations
### Timing of randomization

An important element in designing a TwiCs study is the timing of randomization, which varies according to intervention or treatment under study [48]. Cohort participants can be randomized to the control or intervention arm at one moment in time, which is a feasible approach in a closed or recruiting cohort and is referred to as the 'single-batch sampling approach'. An alternative to the single-batch sampling approach is the 'multiple-batch sampling' approach [48], where a subgroup of cohort participants is randomized at one moment in time. This approach is also feasible for closed or recruiting cohorts and in this approach, the cohort continues to randomize eligible patients who are not allocated yet to the control or intervention arm. Multiple rounds of randomization are conducted within the cohort. This approach was applied in the UMBRELLA FIT trial [39,9,40] and is also adopted in the HONEY trial [37].

　　　　For some interventions, the single and multiple-batch randomizations are not feasible, because screening for trial eligibility and randomization needs to take place within a short timeframe and shortly after diagnosis or after progression or relapse [48]. This means that eligible patients should be randomized as soon as they consented, which makes it impossible to randomize all patients at the same time. This randomization procedure is comparable to the way patients are randomized in standard RCTs. Within a cohort setting, this randomization approach often requires a recruiting cohort and can be applied shortly after the start of the cohort. The latter implies that upon (cancer) diagnosis, patients are invited to participate in a cohort study in which they provide cohort consent and possible consent to be randomized into future RCTs (two-staged informed consent procedure). In case the intervention or treatment needs to be administered also shortly after diagnosis, eligible patients are randomized immediately or very soon after cohort enrollment. In these situations, it is impossible to leave much time between cohort enrollment and the moment patients are randomized into a TwiCs study. This procedure was applied in the RECTAL BOOST trial [32,33,34], where patients provided informed consent for cohort enrollment after being diagnosed with locally advanced rectal cancer and directly after cohort enrollment, patients who consented for randomization into future RCTs were already randomized to the control arm or to the alternative treatment arm. Patients in the control arm received standard chemoradiation and patients randomized to the alternative treatment arm were offered a boost before chemoradiation. By nature of the design, patients in the control arm were not informed about this boost possibility. The same procedure was applied in the VERTICAL trial [43,44,45,46].

When randomization into a TwiCs study starts at the same day or shortly after cohort enrollment, it is inevitable that the 'future' trial is already known by researchers upon the moment that patients sign the two-staged informed consent. This may still lead to selection bias in the trial, which is exactly what one wants to minimize when conducting a TwiCs study. Furthermore, this potential selection bias into the trial brings in another possible risk: selection into the trial may trickle down to selection for cohort enrollment and hence representativeness of the cohort. When a newly diagnosed cancer patient is suited for cohort enrollment, but ineligible for the TwiCs study upon diagnosis, it is highly undesirable to exclude that patient from the cohort. This risk plays a potential role when the TwiCs study investigates the effect of (new) interventions of which it is known that these interventions start shortly after cohort enrollment. The advantages of TwiCs studies over standard RCTs (e.g. fast accrual) should not tempt researchers to start a cohort study for the sake of a clinical trial as this would slowly turn the trial into and RCT following the controversial Zelen design, where patients are randomized before consent is given [49].

**Non-compliance in the alternative treatment arm**
In a TwiCs study, only patients randomized to the alternative treatment arm are asked to provide informed consent *after* randomization (but *before* treatment). As a consequence of this design feature, patients randomized to the alternative treatment are allowed to refuse this treatment after which these patients receive SOC. This will lead to non-compliance in the alternative treatment arm that needs to be considered when calculating the required sample size. Furthermore, as only patients randomized to the alternative treatment arm can refuse treatment *after* randomization, it is crucial to consider how this is accounted for in answering the research question and in determining the effect size, because it is highly unlikely that this refusal of treatment is randomly distributed over study arms (as opposed to standard RCTs). In the remainder of this manuscript, non-compliance is defined as refusal of an alternative treatment or intervention, if offered, after randomization.

Most oncology TwiCs studies presented in Table 1 anticipated on the occurrence of non-compliance in the treatment arm during the design phase by including the expected non-compliance rate in the sample size calculations. However, an issue is that the anticipated non-compliance rate might deviate from the actual non-compliance rate. This was the case for all completed trials presented in Table 1. In the UMBRELLA FIT trial, the anticipated non-compliance rate was 30%, but after 152 patients of the initially required 166 patients were recruited, it turned out that the actual non-compliance rate was 45% [9]. In the RECTAL-BOOST trial, there was an overall non-compliance rate of 27% compared to an expected rate of 20% [33]. In the VERTICAL trial, the assumed non-compliance rate was 10% while the actual rate was 27% [44]. As the TILT trial was a feasibility study, non-compliance rate in the alternative treatment arm was considered a primary outcome measure, but the authors also included failure to complete follow-up in the control arm in the non-compliance definition, which is why in the TILT trial the non-compliance rate definition was different compared to the other trials [30]. The study was considered feasible with respect to the non-compliance rate if that rate was below 10%. Of the 12 randomized patients, one patient in the alternative treatment arm refused the treatment after randomization and one control patient did not complete the follow-up schedule, which indicates that the 10% maximum was exceeded.

These results show that the actual non-compliance rates deviated from the expected non-compliance rate. The non-compliance rate in the treatment arm can be interpreted as a methodological challenge of a TwiCs study that requires careful consideration when defining the research question, the clinical endpoints and the determination the required sample size. In the upcoming subsections, we will discuss the implications of non-compliance for the treatment effect estimate and the statistical power. However, before these aspects are discussed, it is necessary to first clarify which effect is estimated in a TwiCs trial and how this is connected to the research question. This will be the topic of the next subsection.

**Defining the efficacy estimand in a TwiCs study**
For this discussion we consider the guidelines outlined in the ICH E9(R1) draft addendum on "Estimands and Sensitivity Analysis in Clinical Trials" [50]. The estimand of a clinical trial can be defined as the targeted treatment effect that reflects the research question which is given by the research objective. It provides a summary at the level of the population of what the treatment effect would be in the same patients under different treatment options being compared. How the estimand is to be estimated should be specified in advance of the trial and once this is defined, the trial can be designed as such that it is possible to generate a reliable estimation of that treatment effect. For the definition of the estimand in a clinical trial, it is required to anticipate on so-called intercurrent events, which are defined as events that mark a change in the course of treatment and that influence the estimation and interpretation of treatment effects. Intercurrent events need to be addressed *a-priori* when describing the clinical research question of interest. In a TwiCs trial, non-compliance, or refusal of the alternative treatment *after* randomization but *before* started treatment, can be regarded as such an intercurrent event. It is obvious that this phenomenon will alter the interpretation of the treatment effect and should be considered when defining the estimand. More specifically, what is the estimand in a TwiCs study, which effect is of interest (what is the research question?) and how can we estimate that effect? For the remainder of this discussion, we only consider the refusal of an offered alternative treatment or intervention after randomization as known intercurrent event in a TwiCs study and therefore only discuss the implication of that particular event.

First, it is important to assume that non-compliance due to refusal only occurs in the alternative treatment arm, which means that the intercurrent event is dependent on the assigned treatment. It is also assumed that the occurrence of non-compliance will affect the treatment effect indefinitely, that is, once a patient refuses offered treatment, the patient will receive the SOC for the remainder of the trial duration. Finally, it is assumed that the control patients do not have access and will not get the alternative treatment since these patients are not informed about the trial. In other words, there is no contamination in the control group.

*Treatment policy strategy*
The way non-compliance is addressed in the trial defines the research question that a TwiCs study is able to answer. One of the strategies to address the research question described in the ICH E9(R1) draft guidance document is the treatment policy strategy. This basically means that the intercurrent event is taken to be part of the treatment regimen of interest. The treatment effect is then estimated irrespective of the occurrence of an intercurrent event and the estimand is a combined effect of the initial randomized treatment and the treatment modified by the intercurrent event. Adopting a treatment policy approach has become known as the "Intention-to-treat" (ITT) approach; all patients are analyzed 'as randomized' regardless of the occurrence of the intercurrent event. For a TwiCs study, this implies that the non-compliance rate in the alternative treatment group is considered as part of the treatments being compared.

What does this mean for the interpretation of the ITT effect in a TwiCs study? This question can be answered by first taking a closer look at the ITT definition in a standard RCT. The ITT effect in a standard RCT is generally interpreted as the average causal effect (ACE) of the assigned treatment. The ACE measures the difference in the mean outcome between patients assigned to the alternative intervention and patients assigned to SOC. It has been argued that the ACE of a standard RCT is, on average, an unbiased estimate of the population mean effect of the alternative treatment compared to SOC in patients receiving treatment, under the assumption that treatments are randomly assigned to patients, thereby assuming there exists no confounding [51,52]. Assuming that all patients also receive the assigned treatment, we will refer to the ITT effect in a standard RCT as the ACE of received treatment for the remainder of this discussion. Although this technically is not the pure ITT definition (analyzing 'as randomized' regardless of taking up treatment), this nuance is important to emphasize on the difference between the ITT definition of a standard RCT and a TwiCs study. In a TwiCs study, patients are also assigned a treatment, but the difference lies in the fact that patients are offered alternative treatment when assigned to that treatment, where in a standard RCT we expect all patients to receive the assigned treatment. Therefore, to distinguish between the ACE of a standard RCT and a TwiCs study, we refer to received treatment and offered treatment, respectively.

Non-compliance is known to be a methodological problem that can lead to bias in estimating the ACE of received treatment in randomized experiments [53]. In a standard RCT, however, the refusal of treatment happens generally before randomization and hence, these patients do not enter the trial. Furthermore, in expectation, (potential) non-compliance is randomly distributed over the treatment arms in a standard RCT, which will therefore not immediately cause bias in estimating the ACE of received treatment. In fact, only selective non-compliance in a standard RCT might lead to a more-or-less biased ACE of received treatment relative to the population value of the ACE of received treatment.

In contrast, for a TwiCs study, it is already expected beforehand that non-compliance only occurs in the alternative treatment arm *after* randomization; the intercurrent event occurs by nature of the design. As a result, non-compliance is known to be not random (selective non-compliance) and therefore, the treatment effect under the ITT-principle will be diluted when incorporating non-compliant patients. Intuitively, one might argue that non-compliance in a TwiCs leads to a biased ACE of received treatment, but this is incorrect, because a TwiCs study simply adopts a different estimand compared to a standard RCT. As stated earlier, the ITT effect of a TwiCs study is the ACE of offered treatment rather than received treatment. This also means that when we speak of bias in a TwiCs study, it is important to refer to bias in the estimand of a TwiCs study. In the situation where the refusal rate of the alternative intervention in the trial matches that of the population, the ACE in a TwiCs study will provide an unbiased estimate of the population mean effect of the alternative intervention compared to SOC in patients who are offered the alternative intervention compared to patients receiving SOC [52]. The key point here is that a TwiCs study and a standard RCT estimate a different ITT effect (estimand) under the treatment policy strategy and do therefore answer different research questions. Bias in a standard RCT is therefore defined as bias relative to the effect of received treatment, whereas bias in a TwiCs is defined as bias relative to the effect of offered treatment. Consequently, a TwiCs study will not provide a biased estimate of the ACE observed in a standard RCT, what is sometimes falsely claimed (see Section 3.5).

In the UMBRELLA FIT-, the RECTAL BOOST-, and the VERTICAL trial, the primary analysis was done according to the ITT principle. However, as explained above, interpreting the ITT effect of a TwiCs study cannot be separated from the non-compliance rate in the alternative treatment arm. Therefore, the expression of the final results should be stated carefully. For example, in the VERTICAL trial, the interpretation of the results was expressed as: "we found no differences in pain response, pain scores, and global QOL between patients receiving cRT and those (offered to be) treated with SBRT" [p. 363, [44]]. The part between brackets points out that treatment effects represent the effect of offered alternative treatment compared to receiving SOC rather than a comparison of patients receiving two different treatments [52]. The same phrasing with respect to the treatment effect under the ITT principle was adopted when presenting the UMBRELLA FIT trial results. In addition, for the UMBRELLA FIT trial, results were reported for patients offered the alternative intervention as well as those for patients accepting the alternative intervention [40].

Finally, analyzing a TwiCs study according to the treatment policy strategy ensures that the occurrence of the intercurrent event is also of main interest [54], which means that a TwiCs study can be used to gain insight in the acceptability of an alternative treatment. This was recognized in the VERTICAL trial, where this acceptability was explicitly stated when discussing the results [44]. Therefore, acceptability of the alternative treatment could be part of the research question and must be seen as part of the treatment effect [55].

*Principal stratum strategy*

In addition to the treatment policy strategy, the ICH E9(R1) guideline lists four other strategies to address the research question. Each of these strategies approaches a different research question. We will briefly discuss one other strategy that plays a role in the TwiCs setting. This strategy is the principal stratum strategy. In the principal stratum strategy, the intercurrent event is considered a confounding factor when estimating a treatment effect. The essence is that the treatment effect is estimated in a (target) population ("stratum") whose status with respect to the intercurrent event is similar, irrespective of treatment arm. For a TwiCs study, this means that the treatment effect is estimated in a population that is capable and willing to accept the treatment being assigned to. Using different analysis strategies than the ITT approach, an estimate of the treatment effect under perfect compliance can be generated, typically based on causal inference models [56]. An example of such an estimate is the complier average causal effect (CACE), which provides an unbiased treatment effect for patients who comply with the protocol [57]. This definition diverges from the ITT definition in a TwiCs study, which demonstrates that both estimands are concerned with a different research question.

The remaining strategies listed in the ICH E9(R1) may also apply to the TwiCs design, but the treatment policy strategy and the principal stratum strategy have been described in publications of TwiCs trials, which is why we limit the discussion to these two strategies. For a detailed overview on how to define the estimand based on difference strategies with detailed examples, see [54,58,59].

In sum, different research questions can in principle be addressed by a TwiCs study. The research question drives the definition of the estimand(s) of interest in a TwiCs study, which should be defined before start of the study. These definitions will then determine the primary analysis and, importantly, power and sample size assessment. It is important to realize that these different estimands should not be interpreted as alternatives to one another, but merely as ways to answer different research questions.

**Analysis of a TwiCs study**

As discussed in Section 3.3, the effect of the alternative treatment arm compared to control can simply be estimated by comparing the group of patients randomized to the alternative treatment arm to the group of patients randomized to SOC, using an appropriate statistical test. This approach is similar to the primary analysis strategy of most randomized trials. However, the result of this analysis in a TwiCs study should not be interpreted as the ACE observed in a standard RCT, because the non-compliance rate observed in the intervention arm dilutes this effect and should be considered when interpreting the results.

When the interest is focused on the effect of the intervention under compliance (principal stratum strategy), the analysis must be adapted accordingly. In the TwiCs literature, instrumental variable (IV) analyses have been proposed to accomplish this [55,60,61]. These IV analyses use a two-stage least squares method, to account for possible non-compliance in the alternative intervention group [62]. In the first stage, the effect of exposure (actual treatment received) is predicted by the effect of randomization. In the second stage, this information is used to understand how the exposure affects the outcome. Two different IV analyses were proposed by Pate et al. [61] and Candlish et al. [60] to analyze TwiCs studies. In the first IV analysis, a two-stage regression model is applied. In the first stage, the effect of randomization on exposure is estimated using logistic regression, which provides the estimated exposure given the allocated treatment. Subsequently, in the second stage, a regression model for the outcome is fitted using the estimated exposure from the previous logistic regression model as covariate. The effect of the estimated exposure on the outcome provides the estimated treatment effect of interest. The second IV analysis also starts with a logistic regression model predicting exposure by randomization, but now the residual term is calculated as the difference between actual exposure and predicted exposure. In the second regression model, the outcome is modeled as a function of the treatment received and the residuals calculated from the previous logistic regression where the coefficient of treatment received provides the estimated treatment effect.

In two simulation studies, the performance and accuracy of the ITT and IV analysis in analyzing TwiCs study results was investigated [60,61]. The authors reveal that the larger the refusal rate, the more bias found in the ITT effect as expected in a standard RCT, but considering our arguments in Section 3.3, this is rather a logical finding. When acknowledging that a TwiCs study estimates a different ITT effect compared to a standard RCT, it is expected that the ITT effect of a TwiCs study deviates from a (simulated) ITT effect of a standard RCT, but that should not be interpreted as bias. Again, bias in the ITT effect of a TwiCs study should not be seen as bias relative to the ITT effect of a standard RCT, but relative to its own definition. For example, when non-compliance depends on certain patient characteristics (only male participants refuse treatment), we can expect bias in the ACE of offered treatment relative to the population value.

With respect to the completed TwiCs oncology studies (Table 1), only the UMBRELLA FIT trial provided results of an ITT and IV analysis. In addition, in the UMBRELLA FIT trial, another alternative analysis strategy was used, namely a propensity score analysis by comparing intervention accepters to patients in the control group who would have accepted the alternative intervention if offered [40]. This propensity score analysis serves as a sensitivity analysis to the IV analysis, because it is unknown whether intervention refusers are influenced by the offer of the intervention.

**Statistical power**

In general, sample size calculations should be based on the anticipated treatment effect according to the ITT definition. The anticipated ITT effect of a TwiCs study reflects the ITT effect considering non-compliance in the alternative treatment arm (offered treatment) and will therefore, in general, be smaller than the ITT effect in a standard RCT. As a result, required sample sizes for obtaining sufficient power in a TwiCs study are often larger than those of standard RCTs [60,61] .

A critical issue in TwiCs studies is that the expected non-compliance rate may diverge from the actual non-compliance rate, which was the case in the UMBRELLA FIT trial, the RECTAL BOOST trial, and the VERTICAL trial (see Section 3.2). Consequently, the sample size had to be updated during the trial based on the actual non-compliance rate, which was also recommended by Candlish et al. [60]. This may have severe implications when the observational cohort is limited in the number of available patients, which can be the case in a closed cohort [48]. Updating the required sample size is easier in recruiting cohorts. Furthermore, recruiting cohorts have the advantage that the non-compliance rate can be updated after each randomization and the sample size can be adapted until the actual non-compliance rate is reached. It has been recommended to calculate the required sample size under different scenarios during the design stage under various non-compliance assumptions [48,52] or to first perform a pilot study before the actual TwiCs study to obtain insights in the actual refusal rates [9].

As a final note on the sample size we would like to point out that the discussion of the (diluted) ITT effect so far holds for superiority trials. A diluted ITT effect makes it easier to demonstrate non-inferiority or equivalence. In general, the ITT effect in non-inferiority trials is anti-conservative [63]. Therefore, in designing and analyzing TwiCs non-inferiority trials, a per protocol analysis excluding non-compliance should be considered. However, since non-compliance only occurs in the alternative treatment arm, it is unclear how this will affect treatment group balance and hence the interpretation of non-inferiority. To our knowledge, there have been no proposed or conducted non-inferiority TwiCs studies to date.

**Discussion**

This article provides an overview of TwiCs studies conducted in the oncology setting, where the TwiCs design has gained increasing popularity during the last decade, especially in the Netherlands. The rise in initiated and conducted TwiCs studies is related to several drawbacks that are encountered when performing standard RCTs for which the TwiCs study design offers possible solutions. These drawbacks are the risk of uncompleted trials partly due to slow and difficult accrual of patients, limited external validity in standard RCTs and high drop-out rates in the control group. Whether the latter is an issue in standard oncology RCTs can be questioned, since oncology patients already receive the best SOC and study treatment is not commonly available (except for supportive care setting).

The main elements of a TwiCs study are that these are conducted within an observational cohort study, that patients randomized to the SOC receive no information about the trial and that patients randomized to the alternative treatment arm can refuse this treatment after randomization. The advantage of an available observational cohort is that patients can more easily be found and contacted which will improve accrual speed. In two oncology trials, it indeed was concluded accrual was faster compared to standard RCTs [9,48]. Furthermore, another possible advantage of the availability of an observational cohort is that the collection of routine cohort measurements may minimize selection bias if eligibility is based on these measurements and less affected by selection of a physician [9]. At the same time, it is crucial to capture sufficient key demographics and other historical, disease specific variables at baseline upon cohort enrollment and that this information is regularly updated to determine eligibility for future TwiCs studies. Minimization of selection bias will likely improve external validity. Also the fact that control patients are not informed will improve external validity, as this design element more closely resembles clinical practice.

Despite the possible advantages of a TwiCs study, the design is entailed with several methodological challenges that are addressed in this article. One challenge that is discussed here is the timing of randomization and the associated possible risk of selection bias for cohort enrollment. Another important challenge is the anticipated non-compliance (refusal) rate in the alternative treatment arm and how this non-compliance rate affects the definition of the estimand, the related research question, the analysis methods and the sample size calculation. Important to emphasize here is that the ITT effect of a TwiCs study has a different meaning than the ITT effect of a standard RCT. The ITT effect of a TwiCs estimates the ACE of offered treatment, whereas the ITT effect of a standard RCT estimates the ACE of received treatment. Consequently, the two ITT estimates are different by definition and the difference between the two estimates should not be considered bias, but merely as two estimates that represent answers to two different research questions.

An important question related to the ITT definition in a TwiCs study is whether the results of a TwiCs study could be accepted by regulatory authorities for approval of a new treatment. For pivotal trials, a TwiCs study loses some benefits relative to a standard RCT. However, the loss of benefits is not primarily related to the fact that a TwiCs study provides no effect of received treatment, since a more conservative treatment effect estimate, the effect of offered treatment, may not necessarily be an issue for regulatory authorities. For regulatory authorities and reimbursement policies, concerns may be related to the fact that the population of interest for which a treatment had a positive benefit-risk is difficult to determine a-priori when the two treatment arms provide non-matching safety data due to non-random refusal across treatment arms.

The challenges faced in standard RCTs described at the beginning of this article might motivate researchers to conduct single-arm trials where external, historical controls can be used for comparison purposes. However, the lack of randomization in single-arm trials can cause serious bias and confounding and leads to difficulties in quantifying a treatment effect. Although methods have been proposed to adjust for confounding, bias, and imbalance, randomization is the only way to make detection of imbalances possible in the first place [64]. Therefore, although a TwiCs study is faced with methodological challenges and may not be preferred by regulatory authorities over standard RCTs, it can be considered a more suitable alternative compared to single-arm trials, also for regulatory authorities, because patients are randomized in a TwiCs study, thereby reducing sources of bias that are encountered in non-randomized trials. Notably, the choice of design in the oncology setting is dependent upon many factors. For example, in oncology, ethical aspects play an important role in deciding whether randomization is feasible, as does the willingness of patients to be randomized. For a very thorough discussion and detailed overview of perspectives on the use of randomization in oncology trials, see Grayling et al. [65].

A final point of discussion is the validity of the assumptions stated in this article. We assumed that the occurrence of non-compliance (that is, refusal of assigned treatment as intercurrent event) only occurs in the alternative treatment arm, which is a valid assumption by nature of the design. However, this assumption is violated when patients randomized to SOC refuse to be treated with SOC. Refusing SOC is generally not very common in oncology (see for example three studies on the predictors associated with treatment refusal in colon [66], breast [67], and head and neck cancer [68], where small percentages of SOC refusal were reported), but it cannot be ruled out. How to deal with this non-compliance in the control group in a TwiCs study context with respect to the analysis and interpretation of the results will depend on the research question and whether this control non-compliance will be judged problematic. As such, researchers could determine a-priori whether this is likely to occur and then alter, for example, the expected effect size. Also, the assumption of no contamination in the control group may be violated when these patients become aware of the availability of an alternative treatment, either because patients communicate with each other in the waiting room, or because patients might find information online as trials need to be prospectively registered. In such circumstances, control patients might take the initiative in asking their treating physician to receive the alternative treatment. This may cause potential (ethical) dilemmas for trial staff and possible cross-over of control patients when these patients are already randomized to SOC, without knowing it. However, whether this will be a substantial risk should be considered per trial.

## Conclusion

In conclusion, in this article we provided an overview of potential advantages of a TwiCs study compared to standard RCTs and we reflected upon the most important methodological challenges of a TwiCs study, based on experiences in oncology. Researchers in oncology and other areas should carefully consider these methodological challenges when planning to initiate a TwiCs study.

## List of abbreviations

ACE: Average causal effect; ITT: Intention-to-treat; IV: Instrumental variable; PLCRC: Prospective Dutch colorectal cancer cohort; PICNIC: Prospective data collection initiative in colorectal cancer; PRESENT: Prospective evaluation of interventional studies on bone metastases; RCT: Randomized controlled trial; SOC: Standard of care; TwiCs: Trial within cohorts study; UMBRELLA: Utrecht cohort of multiple breast cancer intervention studies and long-term evaluation

## Ethics approval and consent to participate

This manuscript does not involve the use of any animal or human data or tissue.

## Consent for publication

Not applicable.

**Availability of data and materials**
Any materials used during this study are available from author R. Kessels upon request via r.kessels@nki.nl.

**Competing interests**
AM: Scientific advisor COMPASS pathways, paid to institution. MK: advisory role for Nordic Farma, Merck-Serono, Pierre Fabre, Servier. Institutional scientific grants from Bayer, Bristol Myers Squibb, Merck, Personal Genome Diagnostics (PGDx), Pierre Fabre, Roch, Sirtex, Servier. RK and KR have declared to have no competing interests.

**Funding**
No funding was provided for this paper.

**Authors' contributions**
RK, AM, and KR conceptualized the idea. RK drafted the original manuscript. AM, MK, and KR reviewed and provided feedback on all drafts. All authors edited and reviewed the final manuscript.

**Acknowledgements**
None.